\documentclass[aps,prx,twocolumn,showtitle,longbibliography]{revtex4-2}

\usepackage[unicode=true,pdfusetitle,
 bookmarks=true,bookmarksnumbered=false,bookmarksopen=false,
 breaklinks=false,pdfborder={0 0 0},pdfborderstyle={},backref=false,colorlinks=true]
 {hyperref}
 \hypersetup{
 citecolor=blue,
 urlcolor=blue,
 linkcolor=blue}
 
\usepackage{graphicx}
\usepackage{dcolumn}
\usepackage{bm}
\usepackage{stackrel}
\usepackage{amsmath,amsfonts}
\usepackage{nicefrac}
\usepackage{float}
\usepackage{xcolor}
\usepackage[normalem]{ulem} 
\usepackage[final]{pdfpages}

\makeatletter
\AtBeginDocument{\let\LS@rot\@undefined}
\makeatother

\newcommand{\new}[1]{{\color{black} {#1}}}

\begin{document}
\title{Topological protection breakdown: a route to frustrated ferroelectricity}

\author{Ludovica Falsi$^{1, 2}$}
\author{Pablo Villegas$^{2, 3}$}
\email{pablo.villegas@cref.it}
\author{Tommaso Gili$^{4}$}
\author{A. J. Agranat$^{5}$}
\author{E. DelRe$^{1, 6}$}

\affiliation{$^1$Dipartimento di Fisica, Università di Roma “La Sapienza”, 00185 Rome, Italy}
\affiliation{$^2$Enrico Fermi Research Center (CREF), Via Panisperna 89A, 00184, Rome, Italy}
\affiliation{$^3$Instituto Carlos I de F\'isica Te\'orica y Computacional, Universidad de Granada, Granada, Spain}
\affiliation{$^4$Networks Unit, IMT Scuola Alti Studi Lucca, Piazza San Francesco 15, 55100- Lucca, Italy.}
\affiliation{$^5$The Institute of Applied Physics, The Hebrew University, Jerusalem 91904, Israel}
\affiliation{$^6$Institute for Complex Systems, National Research Council, Rome 00185, Italy}

\begin{abstract}
\new{Phases manifesting topological patterns in functional systems, like ferroelectric and ferromagnetic vortex superlattices, can manifest intricate and apparently ungovernable behavior, typical of frustrated non-ergodic states} with high-dimensional energy landscapes. This is also the case for potassium-tantalate-niobate (KTN) crystals. These transparent ferroelectrics manifest remarkable but little-understood metastable domain patterns at optical (micrometer and above) scales near the cubic-to-tetragonal structural phase transition. Here, we formulate the Topological Breakdown Model based on the competition between intrinsic scales of domain-domain collinear and non-collinear interactions associated with polarization-charge screening. The model is able to explain observed KTN mesoscopic domain patterns and phase diagram as a function of temperature and external electric field. Findings include a precise set of sharp and broad percolative transitions that are experimentally verified, validating our model. Our study identifies the central role played by competing topologically protected states, identifying \new{a} fundamental link between topological protection and frustration that supports a hitherto unexplored functional non-ergodic arena.
\end{abstract}




\maketitle

\new{Disordered ferroelectric crystals can manifest history-dependence and frustrated behavior in proximity of their structural phase transitions, features that affect response to various degrees, becoming dominant in so-called relaxor ferroelectrics \cite{Viehland1990,Bokov1997,Bokov2006,Bellaiche2011,Bokov2012,LandauSC}. The regime forms a modelling challenge, as conventional frustration mechanisms, including Griffiths phases and glassy states, fail to fully account for key experimental features  \cite{Bokov2006,Westphal1992}. In a subset of disordered ferroelectrics, including solid-solution potassium-tantalate-niobate (KTN), the complex relaxor-like  behavior \cite{Bokov2006,Bokov2012} coincides with the emergence of domain ordering in the form of ferroelectric superlattices or supercrystals \cite{Pierangeli2016}. These patterns appear connected to built-in striations \cite{Nishinaga2014, Agranat2007} or engineered heterostructures \cite{Stoica2019,Hadiji2021}, coupled to the reduction of volume polarization charge $\nabla \cdot \mathbf{p} \simeq 0$, where $\mathbf{p}$ is the spontaneous polarization field.  The constraint introduces closed-loop or flux-closure conditions, so that energetically preferred structures appear in the form of ferroelectric topological defects \cite{Luk2020, Govinden2023}. These have topological protection, in that each domain in the pattern can be reoriented only at the expense of modifying all the others \cite{Junquera2023}.  The question naturally arises if there is a connection between these topological features and the emergence of complexity-driven frustrated behavior. In fact, the volume charge screening condition  makes $\mathbf{p}$ topologically analogous to a magnetic induction field, suggesting that domain-domain interaction can also be governed, at the mesoscale, by effective interaction terms that are typical of magnetic interaction. At present, no effective domain interaction model has been proposed to take into account of both standard collinear dipole-dipole interaction and effective magnetic-like noncollinear interaction, leaving the connection between closed-flux topologically-protected domain patterns and frustration largely unexplored.}

\new{We here formulate the Topological Breakdown Model (TBM) with the aim of identifying the role of effective non-collinear interaction in determining complex mesoscopic (from several micrometer to hundreds of micrometer) domain pattern behavior in systems supporting ferroelectric superlattices. The model is based on the competition between intrinsic mesoscopic scales through both collinear and non-collinear interactions that emerge geometrically from discrete-inversion-symmetry-breaking and volume charge-screening flux-closure constraints \cite{Kittel1949,Landau1935}. Computational analyses and real-space renormalization group (RG) techniques allow us to describe how ensembles of flux-closure patterns and superlattices react to external stimuli. Predictions of the domain pattern phase diagram versus temperature and versus electric bias field identify a  frustrated  metastable phase that is experimentally confirmed in near-transition KTN:Li using percolation analyses. Our findings link directly topologically protected defects to frustration and expand our ability to design and control ferroelectric response.}

\section{The Topological Protection Breakdown Model (TBM) for Ferroelectrics}



We develop the TBM starting from the basic fact that the elementary cell of ordinary ferroelectric perovskites $(ABO_3)$ above $T_C$ exhibits a cubic symmetry. In the vicinity of $T_C$, the B-ion features an atomic displacement along the fourfold axis, leading to a discrete inversion-symmetry breaking, which induces a distorted electron distribution and spontaneous electric polarization \cite{Rabe2007}.  \new{We note that this picture holds for many disordered perovskite ferroelectrics, such as KTN \cite{Gumennik2011}, but may not apply to the class of relaxor ferroelectrics, such as PMN, where well-define structural transitions are not observed \cite{Eremenko2019,Gehring2009}}.  Phenomenologically, below $T_C$, a finite sample relaxes into a pattern of different ferroelectric domains, mesoscopic regions in which the polarization is parallel to a given crystal axis.  While each domain enucleates independently with a specific macroscopic polarization $\mathbf{p}$, as these grow, regions with different spontaneous polarization meet, fusing together or forming topological defects in the form of domain walls. Based on Maxwell's electrostatic equations, domains arrange to allow the screening of volume polarization charge, i.e., the minimization of $\nabla \cdot \mathbf{p}$ \cite{Muench2019}. This leads to two types of defects (see Fig.\ref{Sketch}(a)), type i) walls which separate two domains with anti-parallel polarization with $\mathbf{p}_i=-\mathbf{p}_j$  and type ii) walls that form between orthogonal polarizations $\mathbf{p}_i \cdot \mathbf{p}_j=0$. For a given starting close-packed multi-domain distribution, dynamics appear driven by domain dipole-dipole interaction with the constraint that $\nabla \cdot \mathbf{p} \simeq 0$ on the walls.  While dipole-dipole-like interaction, dominant across type i) defects, can be described through the paradigmatic Heisenberg model $(\mathcal{H}_{H}\propto J_{ij}\mathbf{p}_i\cdot\mathbf{p}_j)$, the interaction across type ii) walls is governed by polarization charge screening, i.e., by non-collinear interactions.

Peculiar phenomena involving non-collinear interaction have been discussed in magnetic systems in the form of magnetic topological defects \cite{Yoshimura2016}, spin-orbit torques \cite{Ishikuro2019}, or magnetically driven ferroelectricity \cite{Dong2015}, and are associated to spin-orbit coupling \cite{Takeuchi2021}. In these, the so-called Dzyaloshinskii-Moriya interaction (DMI) \cite{DM1958, DM1960} underlies many chiral topological magnetic structures, such as skyrmions, and leads to a strong impact on spin dynamics \cite{Yang2023}. However, its electric counterpart has rarely been considered. Still, recent first-principles simulations of perovskites \cite{DM} have demonstrated that electric DMI might originate from the existence of oxygen octahedral tiltings in combination with symmetry arguments \cite{DM}. This amounts to the counterpart of the magnetic spin–current model \cite{Katsura2005}. Recent experimental results have shown DMI-induced polar vortices in single thin ferroelectric films \cite{Rusu2022}. In turn, DMI is rather weak in many material systems and is typically accompanied by a much stronger Heisenberg exchange coupling. The empirical relevance of DMI to relaxor ferroelectrics is far from established. We model non-collinear interactions by introducing an energetic term $\mathcal{H}_{D}\propto\mathbf{D}_{ij}\cdot(\mathbf{p}_i\times\mathbf{p}_j)$, that is the analogous of the Dzyaloshinskii–Moriya interaction (DMI) encountered in ferromagnetic systems \cite{Yang2023}. As recently discussed for ferroelectrics \cite{DM}, this type of interaction can arise from specific microscopic mechanisms, such as oxygen octahedral tiltings or anisotropy at mesoscopic scales \cite{DM, Chen2022}. In the TBM, it arises from the charge-screening constraint.

\new{We want to emphasize that the term "breakdown" refers here specifically to the loss of topological protection associated with flux-closure configurations. These configurations, stabilized by the local constraint $\nabla \cdot \mathbf{p} =0$, are dynamically robust and can only be destroyed via coordinated, nonlocal domain rearrangements, thus functionally mimicking topological defects in low-dimensional systems. The breakdown occurs when thermal fluctuations or applied external fields induce a transition from vortex-protected states to states with extended or reoriented domains, inducing the collapse of this protection mechanism.}

The Hamiltonian can be thus written as $\mathcal{H}=\mathcal{H}_{H}+\mathcal{H}_{D}+\mathcal{H}_e$, with $\mathcal{H}_{H}$ representing the Heisenberg interaction, $\mathcal{H}_{D}$ the DMI interaction, and $\mathcal{H}_{e}$ the electrical alignment to the external field, namely,
\begin{equation}
\mathcal{H}=-\underset{\langle i,j\rangle}{\sum}J_{ij}\mathbf{p}_{i}\cdot\mathbf{p}_{j}+\underset{\langle i,j\rangle}{\sum}\mathbf{D}_{ij}\cdot(\mathbf{p}_{i}\times\mathbf{p}_{j})
-\sum_{i}\mathbf{E}\cdot\mathbf{p}_{i}
\label{Hamiltonian}
\end{equation}

where $\mathbf{p}_i$ is the n-component spontaneous polarization distribution of the domain in the "ith" site with discrete values \new{$\mathbf{p}_i=(\pm p_x, \pm p_y)$, $J_{ij}=JA_{ij}$ are the elements of the $N\times N$ adjacency matrix $\hat A$}, with $J$ the global coupling strength, and $\mathbf{D}_{ij}\propto\mathbf{p}\times\mathbf{e}_{ij}$ where $\mathbf{e}_{ij}$ is the unit vector pointing from i-th site to j-th site and $\mathbf{p}$ the polarization vector \cite{Katsura2005}. The last term corresponds to an applied external field and is described, as usual, \new{as $\mathcal{H}_e=-\sum_i\mathbf{E}\cdot\mathbf{p}_i$.}


A first order parameter is taken to be the modulus $\left|\textbf{p}\right|_i$ of the polarization in each direction $P=\frac{1}{N}\sum_i \left|\textbf{p}\right|_i$, where $i$ represents the sum over nodes, and $|p|=\sqrt{p_x^2+p_y^2}$. A second order parameter is taken to be $v=\frac{1}{2N}\underset{\langle i,j\rangle}{\sum}\frac{|\mathbf{p}_i\times\mathbf{p}_j|}{|\mathbf{p}_i||\mathbf{p}_j|}$, this quantifying emergent rotational order in the system (i.e., vortex-like structures). This is an angular order parameter that captures the angular ordering of the system. Note that, unlike low-dimensional systems undergoing a continuous symmetry breaking, where the emergence of topological defects in the form of vortices is expected as a direct consequence of the Mermin-Wagner theorem  \cite{Cassi1992, Mermin1966}, for the present discrete symmetry case, vortex-like patterns are a direct consequence of the DMI.


Dipole-dipole-like interaction is analyzed using the lattice illustrated in Fig.\ref{Sketch}(b),  a 2D lattice  based on a specific underlying mesoscopic arrangement, known as Supercrystals (SCs), observed in a large variety of ferroelectrics \cite{Stoica2019,Hadiji2021,Dai2022} and other functional materials \cite{Fernandez2022}.  Here, the elementary triangular building block encompasses both type i) and type ii) walls with their protected flux-closure patterns, while the resulting structural geometry  allows  a non-vanishing DMI interaction, excluded in structures that have bond-inversion symmetry (such as square lattices \cite{Ferrari2023}, see also Supplemental Material (SM) \cite{SM}).

\section{Results}

\subsection{SC-tiling Intrinsic Scales}
The intrinsic scales of the SC-tiling lattice are analyzed using the recent framework proposed in \cite{LRG, PRR, PRL-SI} (see Appendix \ref{StatphysApp} and SM \cite{SM}).
The spectral analysis of the SC lattice is based on the time-evolution operator $e^{-\tau\hat L}$ of the diffusion or heat equation, where  $\hat L=\hat D-\hat A$ is the Laplacian operator, $\hat A$ the adjacency matrix, and $\hat D$ the diagonal degree matrix \cite{PRR}. Hence, one can define the Laplacian density matrix, $\hat \rho(\tau)=\frac{e^{-\tau \hat L}}{Tr (e^{-\tau \hat L})}$ (see also Appendix \ref{StatphysApp}) to analyze the so-called lattice "heat capacity" \cite{PRR}  as,
$C(\tau)\equiv -\frac{dS}{d \log \tau}$ which describes the rate of information acquired about the network or lattice structure during diffusion dynamics at scale $\tau$. In analogy with statistical physics, peaks of $C$ are associated with structural phase transitions. Thus, we can analyze $C$ at varying $\tau$ to investigate the lattice multi-scale organization.

\begin{figure}[H]
    \centering
    \includegraphics[width=1\columnwidth]{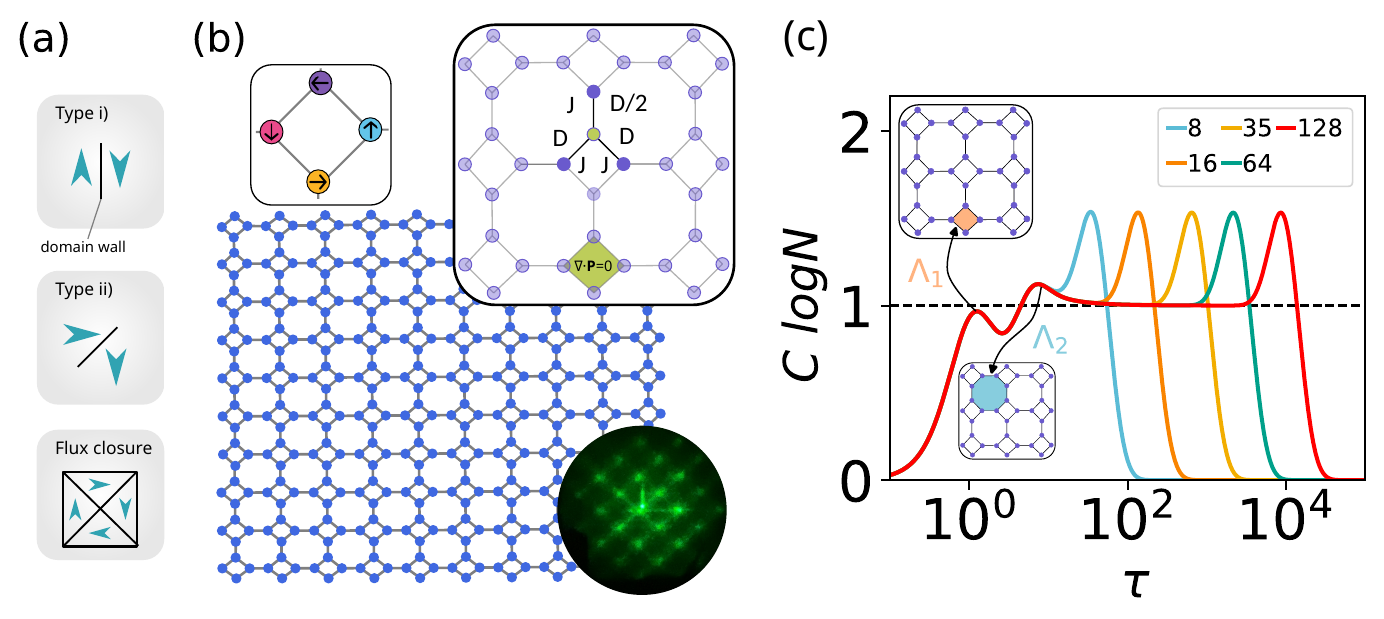}
    \caption{\textbf{(a)} Different types of defects: type i) and type ii), and a flux-closure cell. \textbf{(b)} SC-tiling lattice and sketch of the energetic constants. We highlight the anisotropy in the Dzyaloshinskii–Moriya interaction. The upper left insets show a microscopic vortex, and the lower inset is an experimental far-field imaging of a SC state (right, see below). \textbf{(c)} Specific heat ($C$), versus the temporal resolution parameter, $\tau$, of 2D SC lattices of different sides $L$ (see legend, $N=4L^2$). Insets show the two possible ultraviolet cut-offs, $\Lambda_1$ and $\Lambda_2$, the smallest scales showing translational invariance. The black dashed line highlights the plateau at $C=1$.}
    \label{Sketch}
\end{figure}

Figure \ref{Sketch}(c) shows the so-called specific heat ($C$, see Appendix \ref{StatphysApp}) of the SC structure. Along with the plateau that reflects the 2D nature of the lattice, as $C=\frac{d}{2}$ for scale-invariant architectures \cite{PRL-SI}, $C$ presents two possible small-scale modes or ultraviolet cut-offs (in the jargon of the Renormalization Group), $\Lambda_1$ and $\Lambda_2$, independent of the system size, plus a third peak at larger values of $\tau$ reflecting the whole lattice scale. Compared to a regular 2D lattice that only presents two peaks, here, the extra peak plays, as discussed below, a key role in determining, altering, and disrupting the phases and phase transitions of the system. \new{The presence of multiple characteristic topological scales ($\Lambda_1$ and $\Lambda_2$), each associated with a different type of interaction (Heisenberg vs. DM-like), leads to frustration:  if the system cannot simultaneously minimize both energy terms globally. As a consequence, domains settle into configurations that are locally stable but globally incompatible -- a hallmark of frustrated systems. This manifests in metastability, slow dynamics, and history-dependent responses (see below).}


\begin{figure*}[!hbtp]
    \centering
    \includegraphics[width=2.0\columnwidth]{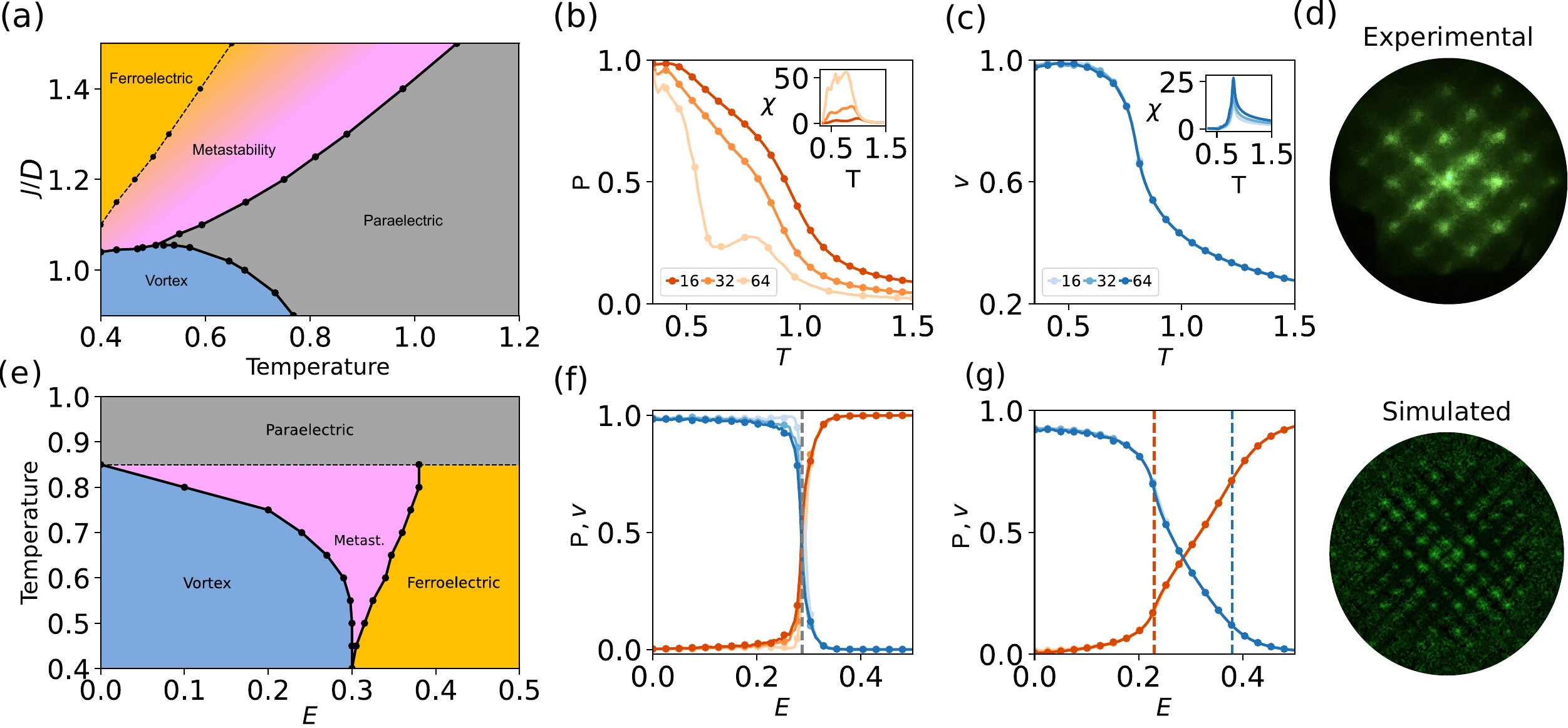}
    \caption{\textbf{Topological symmetry-breaking. (a)} Temperature phase diagram of Eq.\eqref{Hamiltonian} computed using direct simulations on a SC lattice with $N=1024$ nodes ($J=1$). Polarized states are computationally detected with the polarization order parameter, $P$, while vortex phases are detected through the vortex order parameter $v$. The metastable phase corresponds to the broad region of divergent polarization susceptibility $\chi_P=N\sigma^2_P$. \textbf{(b)-(c)} Polarization ($P$) and vortex ($v$) order parameters versus temperature for different lattice sizes (see legend, $N=4L^2$) for (b) $J/D=1.4$ and (c) $J/D=0.8$. Inset: Rescaled temporal variance $\chi_P=N\sigma^2_P$ versus temperature, $T$ for different system sizes. The metastable phase also presents characteristic temporal oscillations in the global system polarization (see Appendix \ref{TempOsc}).
    \textbf{(d)} Top: Far-field imaging of a SC structure at $T$= $T_C$-2K. Bottom: Computational 2D Fourier transform of an ordered vortex phase ($J/D=0.8,~T=0.4,~N=1024)$.
    \textbf{(e)} Temperature phase diagram of the model versus \new{electric field $E$} for $J/D=0.8$ and $N=1024$ nodes. The breakup of the vortex phase translates into the emergence of a metastable phase where the ferroelectric and vortex domains naturally coexist. \textbf{(f)-(g)} Polarization ($P$) and vortex ($v$) order parameters as a function of the electric field for different lattice sizes (as in (b) and (c)) for (f) $T=0.4$ and (g) $T=0.7$. The peak in the system susceptibility is indicated here as a vertical dashed line. Note how the critical field at low temperatures splits into two different phase transitions when T increases, with naturally coexisting vortices and ferroelectric domains. All curves have been averaged over $10^2-10^3$ independent realizations.}
    \label{PhaseD}
\end{figure*}

\subsection{Phase Diagram}
Figure \ref{PhaseD}(a) shows the temperature phase diagram of the system as a function of the ratio between the dipole-dipole alignment interaction, $J$, and the DMI, $D$. Results are achieved through extensive Monte Carlo simulations of the TBM, running Eq.~\eqref{Hamiltonian} on top of the  SC-tiling. In order to reach zero local charge density, the constraint $\nabla \cdot \mathbf{p}_{\Lambda_1} = 0$ is imposed as the energy is minimized (see Fig. \ref{Sketch}(b)), centering it on each "plaquette" (the elementary protected cell $\Lambda_1$) to avoid nonphysical situations (see SM \cite{SM} for case examples of microscopic configurations, \new{resembling the ice rule in spin-ice systems \cite{ICE})}. Inspection of the phase diagram reveals two stable collective dynamical regimes: vortices and ferroelectric phases. In particular, the microscopic scale $\Lambda_1$ now supports the existence of an ordered domain mosaic of 2D protected vortex cores. Different types of bifurcation lines separate the phases of the system. In particular, for low temperatures, increasing the ratio $J/D$, a first-order phase transition separates the vortex phase from the ferroelectric one. Conversely, a second-order phase transition separates the vortex phase from the paraelectric one. Interestingly, a new phase emerges between the ferroelectric and paraelectric phases: a metastable, \new{or frustrated}, phase where transient vortices prevent the system from reaching a \new{global order as in the ferroelectric or vortex phases}, sequentially changing the full system orientation (see \new{Appendix \ref{TempOsc}). }


Figures \ref{PhaseD}(b) and (c) show order parameters as a function of the temperature for the polarization, $P$, and vortex, $v$, for two different values of the ratio $J/D$. Note that $P$ and $v$ correctly detect the stable ordered phases.  Insets show the anomalous temporal fluctuations $\sigma^2_P$ and $\sigma^2_v$, which exhibit a pronounced peak located at the ($N$-dependent) transition point between the different phases. In particular, the broad divergence of $\sigma^2_P$ \new{is a hallmark of frustrated phases in statistical mechanics supporting the picture of broad critical-like regions \cite{Villegas2, MAM-GP}, and give us a marker for the emergence of a new intermediate phase between the ferroelectric and the paraelectric ones (see also Appendix \ref{TempOsc}). }
\begin{figure*}[hbtp]
    \centering
    \includegraphics[width=1.5\columnwidth]{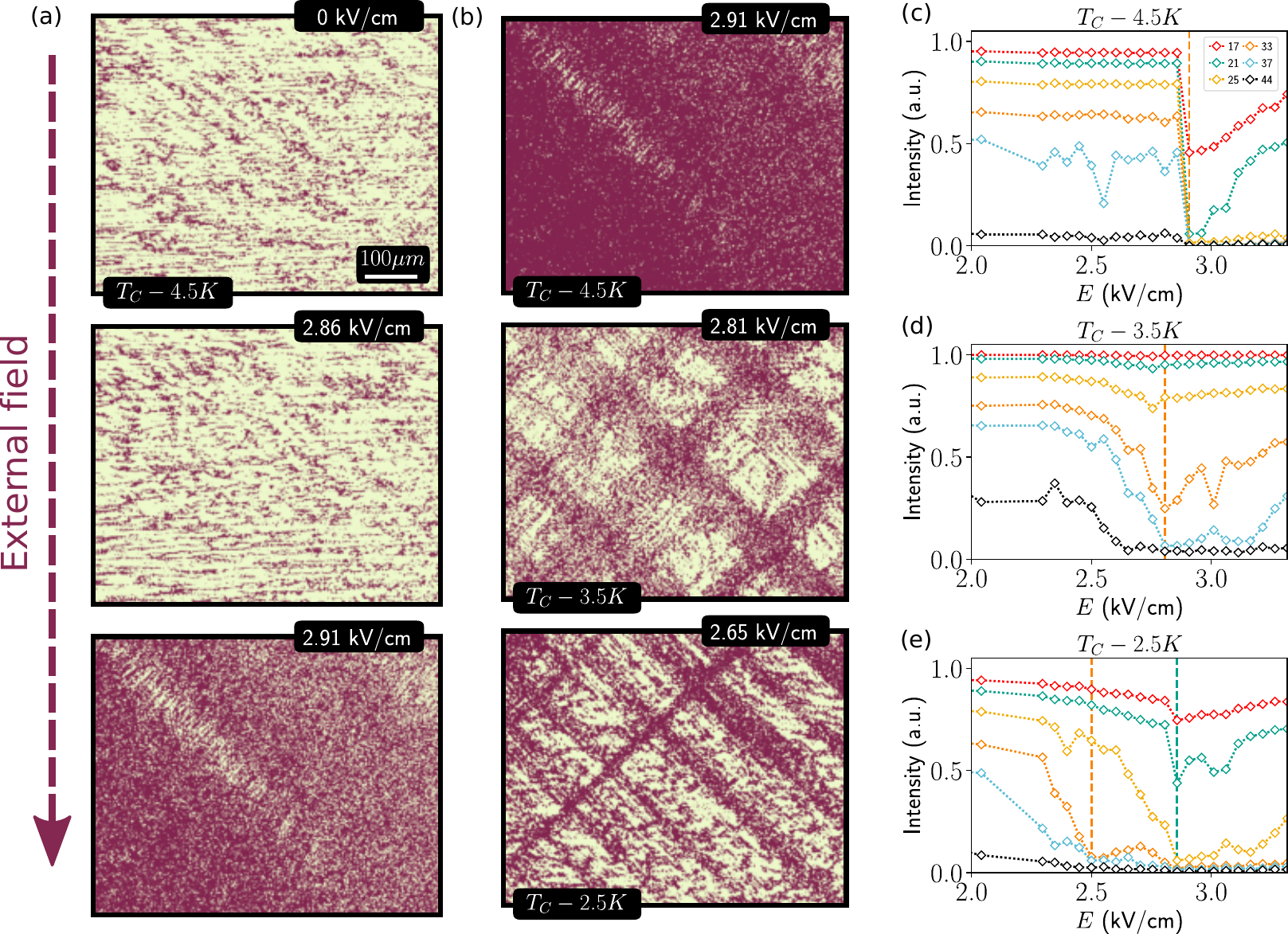}
    \caption{ \textbf{Direct imaging of SC ferroelectric clusters.} Binarized crossed-polarizer transmission microscopy images for \textbf{(a)} $T_C$-4.5K and increasing electric field (0 kV/cm, 2.86 kV/cm, and 2.91 kV/cm, respectively, using $\theta=25$). Light transmission shows an abrupt transition at $E_C\approx$2.9 kV/cm.  \textbf{(b)} Various conditions of temperature
and bias electric field to enhance the critical point for the first two temperatures and the metastable region for $T_C$-2.5K (see images, with $\theta=37$, $37$, and $35$, respectively). \textbf{(c-e)} Light intensity versus applied electric field ($E$) for three different temperatures and different threshold values ($\theta$, see legend): \textbf{(c)} $T_C-4.5K$. An abrupt phase transition is present at a field $E_C\approx2.9$ kV/cm for a wide range of $\theta$ values. \textbf{(d)} $T_C-3.5K$. The abrupt transition changes its nature to a continuous one with $E_C\approx$ 2.75 kV/cm and \textbf{(e)} $T_C-2.5K$. The system exhibits two different phase transitions at different critical fields, $E_C\approx$2.5 kV/cm and $E_C\approx$ 2.86 kV/cm (see also Appendix \ref{PerAnalysis}).}
    \label{RealPerc}
\end{figure*}

Findings are compared to experiments in SCs observed in KTN:Li crystals (see \new{also Appendices \ref{ExperimentalApp} and \ref{MaterialApp}).} This material, which led to the first observation of wave replica-symmetry-breaking \cite{Pierangeli2017}, has been developed by several groups \cite{LiLi2020,Wu2022,Wu2_2022,chang2013} as a lead-free optically transparent alternative to other commonly used perovskites \cite{Noheda1999}. Moreover, compared to commercially available crystals, such as BTO \cite{Fernandez2022} and SBN \cite{shvartsman2012}, it has tunable properties with a room-temperature ferroelectric transition \cite{GR,SHG,Pierangeli2016}. Fourier computational analysis of the vortex phase (see also SM \cite{SM}) is found to match observed optical far-field results, \new{as shown in Fig.\ref{PhaseD}(d)}, in agreement with the previous SC models based on an interlocked regular structure of spontaneous polarization \cite{LandauSC,Falsi2023,SHG}. The TBM model then provides a feasible explanation to why the SC phase appears only in some specific materials, as the internal structure must be fine-tuned to present a specific $J/D$ ratio.

\subsection{The effect of a bias electric field}
Response to a bias electric field is analyzed theoretically by adding the \new{term $\mathcal{H}_e=-\sum_i\mathbf{E}\cdot\mathbf{p}_i$  in Eq. \eqref{Hamiltonian}.}  As reported in Fig.~\ref{PhaseD}(e), the temperature versus electric field phase diagram presents a rich phenomenology. At low temperatures, the vortex phase is separated from the ferroelectric phase by a first-order phase transition, while the critical temperature $T_p$ gives rise to the paraelectric phase (dashed lines). Far below this temperature, a different \new{metastable or frustrated} phase emerges in contrast to the previous case. As before, the order parameters ($P$ and $v$) characterize the phase transitions as a function of the external field, as shown in Figs.~\ref{PhaseD}(f) and (g).




Cluster patterning is directly observed using laser light cross-polarizer transmission microscopy \cite{PercolationSC} (see Appendix \ref{ExperimentalApp} for a detailed description of the experimental setup). Specifically, light propagating in the SC structure becomes depolarized and hence is able to pass through the crossed polarizers \cite{Ferraro2017}. Regions where no transmission is observed identify where the SC is either not formed or has a defect, such as in proximity of a domain wall.
Basic phenomenology and analysis are reported in Fig.\ref{RealPerc}, showing the transition from light transmission, partial, to no light transmission as the SC structure is caused to break down. Increasing the bias field for a sample cooled to $T_C-4.5K$,  at $E_c$ = 2.91 kV/cm the crossed polarizer image suddenly becomes opaque, indicating SC breakup and an ensuing overall depolarized transmitted beam (see Fig.\ref{RealPerc}(a)). Polarization transmission images have been selected to show the percolative nature of the observed phenomena at three different temperatures below $T_C$ = 294K, and for different electric field values $E$, as shown in Fig.\ref{RealPerc}(b). For temperatures up to $T_C-3.5K$, the SC no longer manifests sudden polarization changes. Transmission starts to decrease along particular directions, oriented at 45 degrees relative to the crystal principal axes. Bias-induced distortions are found to predominantly affect the SC structure along these inclined paths, at the critical field value $E_c$ = 2.81 kV/cm. Despite this, light transmission remains evident in the overall image. A different behavior is found at $T_C-2.5K$. Specifically, while analogous SC distortions appear along inclined paths, but at a lower field, $E_c$=2.5 kV/cm, as the bias is increased, the darker regions are observed to expand along these specific directions until they cover the entire transmitted image at $E_c$= 2.86 kV/cm (see Appendix \ref{PerAnalysis}), as seen at $T_C-4.5K$.

The phase transition can be characterized through the usual percolation order parameter $P_\infty=n_{\infty}/N$, where $n_{\infty}$ is the number of pixels belonging to the largest cluster and $N$ is the total number of pixels in the processed image \cite{PercolationSC}. The images are binarized by applying an arbitrary threshold intensity, $\theta$. Each pixel in the image is classified into one of two states, i.e., one or zero, based on whether its intensity is below or above the threshold, respectively. Hence, the largest cluster of light corresponds to the intensity in Fig.\ref{RealPerc}, with the electric field E as the control parameter, showing a percolation phase transition at some critical value $E_{c}$. We note that the proper identification of the translational invariant "building blocks" justifies the dynamical analysis of the system on a coarse-grained scale that can potentially alter the system universality class, as found when the system is biased with an electric field.

Figs.\ref{RealPerc}(c)-(e) show the percolation phase transition for all three selected temperatures below $T_C$ and different values of $\theta$. Experimental results are fully compatible with numerical results reported in Fig.\ref{PhaseD}(e)-(g). For temperatures deep into the ferroelectric phase, $T_C-4.5K$, and $T_C-3.5K$, the system shows a first-order phase transition that splits into two phase transitions for $T_C-2.5K$ (at $E_c$=2.5 and $E_c$=2.86, respectively). Besides, the nature of the phase transitions changes between Fig.\ref{RealPerc}(c) and Fig.\ref{RealPerc}(d), the latter being more likely in the vicinity of some triple point. Note also how the original first-order phase transition (orange dashed lines in Fig.\ref{PhaseD}) shifts towards lower fields as expected from our theoretical framework.

To establish whether the TBM actually supports a percolative phase transition, we analyzed each unit belonging to the microscopic scale, $\Lambda_1$, on a coarser scale: each block-dipole now represents a new node that either allows light to pass ($\ell=1$) or does not ($\ell=0$).  This embodies the idea that for the KTN:Li, only the SC vortex phase allows complete light transmission. The giant cluster $P_\infty$ characterizes the phase transition, together with the susceptibility $\chi(h)=\frac{\sum_{S}S^{2}P(S,h)}{\sum_{S}SP(S,h)}$. Here, the sum runs over all possible cluster sizes $S$ for a given field $h$ in the system, being $P(S,h)$ the cluster size probability distribution, and discarding $P_{\infty}$, if it exists.

\begin{figure}[hbpt]
    \includegraphics[width=1.0\columnwidth]{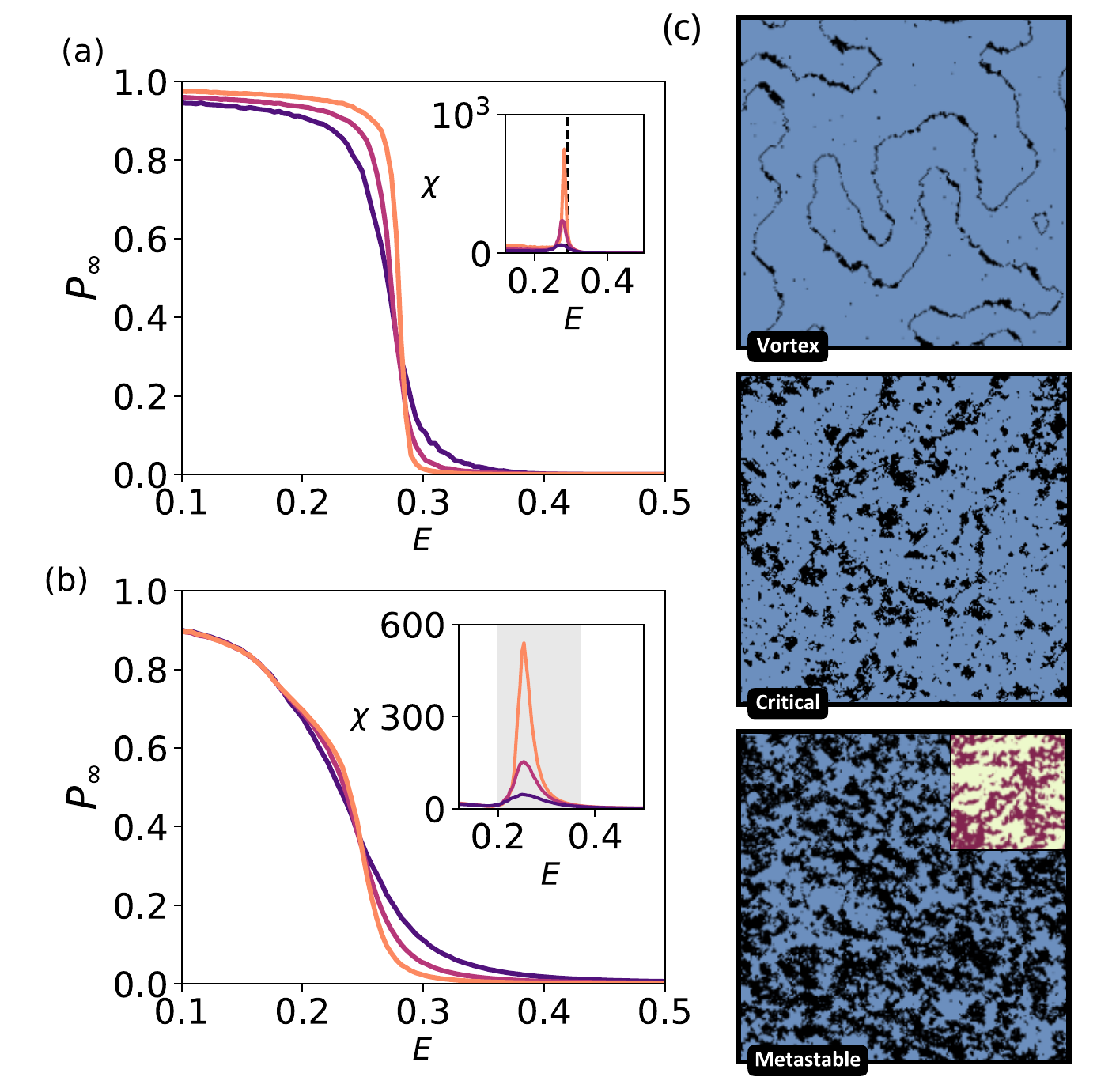}
    \caption{\textbf{Percolation phase transition. (a-b)} $P_\infty$ as a function of the applied field, $E$, for different system sizes (in the new square lattice of side $L$ resulting from the coarse-graining process, see legend) for: \textbf{(a)} $T=0.4$  Inset:  $\chi$ versus $E$. Note the divergence at a specific point as the system size increases. \textbf{(b)} $T=0.75$.  Inset:  $\chi$ versus $E$. Note the divergence of the entire metastable region as the system size increases.  \textbf{(c)} Configurations of simulated light transmission images. From up to down, images show a Vortex state ($T=0.4,~E=0.2$), the bifurcation line to the metastable phase ($T=0.5,~E=0.29$), and the metastable phase ($T=0.5,~E=0.3$). For the sake of comparison, the inset shows a zoomed region of experimental light transmission at $T_C$-2.5K and $E=2.65$ kV/cm.}
    \label{PercSim}
\end{figure}

\new{Figure \ref{PercSim}(a)} shows the percolation phase transition at low temperatures, which confirms that an abrupt phase transition occurs at a specific critical field, $E_c$. Fig. \ref{PercSim}(b) illustrates the behavior in the metastable phase, shedding light on two key properties: (i) the phase transition in the metastable region is found to be smoother, and (ii) susceptibility diverges throughout the entire region. Fig. \ref{PercSim}(c) shows different states when the external field increases. At low temperatures, labyrinthine-like structures form where ferroelectric domains start to aggregate, giving rise to a giant cluster that spans the entire system in some critical field $E_C$. This leads to the \new{frustrated} phase, where both the vortex and ferroelectric mesoscopic domains coexist in an interlocked way, forming fractal patterns (see Fig. \ref{PercSim}(c)). 

\begin{figure}[hbtp]
    \includegraphics[width=1\columnwidth]{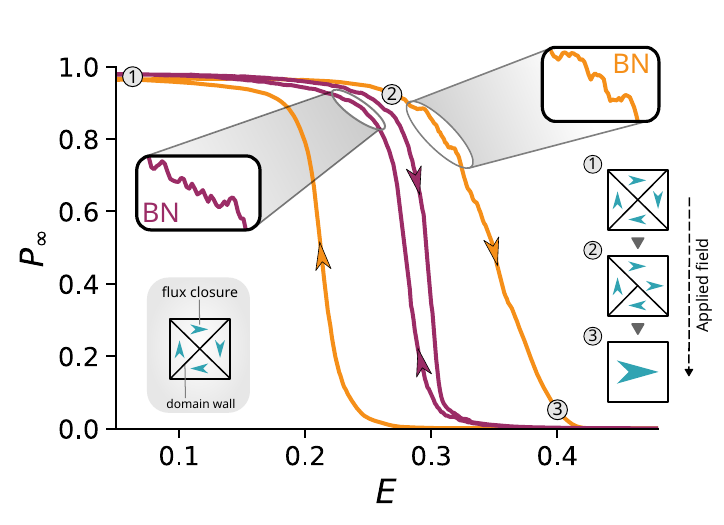}
    \caption{\textbf{Slim loop hysteresis.} $P_\infty$ as a function of \new{the applied field, $E$,} for different temperatures, $T=0.4$ (orange line) and $T=0.6$ (violet line) for a lattice with  $L=64$ coarse-grained units. The wide hysteretic behavior at low temperatures, characteristic of the first-order phase transition, becomes a slim loop hysteresis cycle in the intermediate phase. The break of domain walls on pinned sites originates the observed jumps in the hysteretic cycle, that are shown in the zoomed insets. This gives rise to the well-known Barkhausen noise.}
    \label{Slim}
\end{figure}

Macroscopic response is analyzed in Fig. \ref{Slim}, where the behavior of $P_\infty$ versus applied field is reported. The orange loop  refers to temperature where \new{frustration} is absent, leading to a strongly hysteretic reponse function typical of a first-order transition.  In contrast, a slim loop (violet loop) emerges in the metastable phase, a signature of negligible hysteresis and marked nonlinearity.  Characteristic discrete jumps in polarization (see the zoomed inset in Fig. \ref{Slim}) are caused by the depinning of finite domains, the so-called Barkhausen jumps in hysteretic loops \cite{Colaiori2008, Westphal1992}.

\section{Discussion}

\new{A knot in a string is a remarkably stable physical form of information storage since it is topologically protected, that is, it can only be erased by untangling it \cite{Atiyah1990, Kauffman2001, Rolfsen2003, Turner1996}. 
Knot-like structures emerge naturally in the form of topological defects in a wide variety of materials \cite{Faddeev1997}, such as liquid-crystals \cite{Tkalec2011}, fluids \cite{Kleckner2013}, ferromagnets \cite{Sutcliffe2017, Zheng2023}, and ferroelectrics \cite{Luk2020, Govinden2023}.} This protection is responsible for the stability of intricate low-dimensional topological polar structures (e.g., quadrant domains \cite{Kim2018}, polar flux closures \cite{Kittel1949,Tang2015,Li2020}, and vortices \cite{Seidel2016}) that, in recent years, have attracted significant attention\cite{Stoica2019,Hadiji2021,Bellaiche2004}. Analogous topologically protected patterns naturally occur in a wide variety of systems \cite{Faddeev1997}, such as liquid-crystals \cite{Tkalec2011}, fluids \cite{Kleckner2013}, and ferromagnets \cite{Sutcliffe2017, Zheng2023}.  Among these, topologically protected ferromagnetic skyrmions \cite{Fert2013} have found applications in electromechanical and spintronic devices for information storage \cite{Scott2007} and logic \cite{Du2018}. 

At present, the potential of topological defects as noise-resistant memory and processing elements is largely unexplored, especially because their behavior appears intimately connected with an intriguing but little-understood frustration in the hosting material \cite{Wild2017,Junquera2023,Nahas2016}. In some respects, frustration likens functional materials, such as ferroelectric crystals, to the large class of information-driven phenomena, such as memory effects \cite{Amit1985}, protein-folding mechanisms \cite{Goldstein1992}, brain dynamics \cite{MAM-GP,Villegas2014}, and slow relaxation dynamics \cite{Tetsuaki2017}.   The driving mechanism for frustration is commonly thought to be an underlying disorder that can, for example, influence a conventional phase transition by introducing competing interaction terms between multiple components and lead to frustration \cite{Bellaiche2011}. In ferroelectrics, complex behavior is typically attributed to alterations in what would normally be considered a conventional paraelectric-ferroelectric structural phase transition caused by built-in compositional disorder. Disordered ferroelectrics can be modelled as a system dominated by mesoscopic polar nanoregions (PNRs), regions originally thought to have randomly distributed dipole moments driven by exchange interaction \cite{Cross1994, Burns1983, Westphal1992}, that have more recently been shown to manifest a hierarchical structure with no sudden boundaries \cite{Eremenko2019}. Complex behavior could then be explained as an analog of Griffiths phases theoretically predicted long ago (but not yet experimentally found) for dilute ferromagnetic systems \cite{Timonin}. This approach has been demonstrated to be insufficient at temperatures lower than the Curie temperature ($T_C$) because it fails to consider PNR-interaction \cite{Bokov1997}. The existence of a glassy state has also been proposed as an alternative explanation  \cite{Sherrington2015, Viehland1990}. Glassy approaches, in turn, encounter difficulties in describing some of the basic observed features, such as the occurrence of Barkhausen jumps during the poling process, jumps that are incompatible with the continuous and monotonous glassy reorientation of dipoles on sub-micrometer length scales \cite{Westphal1992}. A complete understanding of emergent relaxor behavior in disordered ferroelectrics remains an open problem \cite{Zheng2024}. 

Our study suggests an altogether different picture. Here, built-in disorder is relegated to generating the SC-tiling and plays no direct role in the TBM.  Yet the approach is able to reproduce the signature and hereto little-understood complex and non-ergodic behavior of disordered ferroelectrics like KTN.

For one, the percolative transition typical of relaxor-like perovskites \cite{PercolationSC, Nahas2016, Prosandeev2013}, is predicted by the TBM at low temperatures (below $T_c$) as an abrupt phase transition that splits into two critical electric fields with an intermediate behavior that resembles that of \new{the new frustrated phase presented here (as discussed in Figs.~\ref{RealPerc} and \ref{PercSim}).} \new{We note that while the cubic-to-tetragonal transition model used here is a useful paradigm for many perovskite ferroelectrics, other systems like PMN and BTO may exhibit more complex or distinct symmetry-breaking behavior. Nonetheless, our model captures the general interplay between discrete inversion-symmetry breaking and charge-screening-driven topology, which remains relevant across a wide class of materials.} Moreover, as discussed in Fig.~\ref{Slim}, the TBM is able to describe the anomalous hysteretic "slim-loop" behavior observed in relaxor behavior, along with the characteristic Barkhausen jumps \cite{Westphal1992}. Our results indicate that complexity-driven phenomenology originates from the competition between the two intrinsically different scales ($\Lambda_1$ and $\Lambda_2$)  that arise not out of disorder but geometrically, from the interaction terms of Eq.(\ref{Hamiltonian}), as discussed in Fig.\ref{Sketch}, leading to local \new{energetic incompatibilities} that persist even in the thermodynamic limit.

\new{Hence, the interplay between the Heisenberg-like and DMI-like terms naturally introduces frustration into the system. These two interactions, effective at different spatial scales, cannot be simultaneously minimized. We want to emphasize that other types of interactions, grounded in different topological scales as supported by the SC lattice, may also lead to similar effects.  As a result, the system is unable to achieve a globally ordered state, leading to the emergence of metastable regimes characterized by local competition between polarized and vortex domains. This frustration is intrinsic to the TBM and forms the basis for the rich phase behavior discussed here.}

The idea that exotic states with degenerate energy surfaces and broad critical phenomena can arise from the breaking of a regular structure of topologically protected states, i.e., the topological symmetry-breaking transition discussed in Figs.\ref{RealPerc} and \ref{PercSim}, presents a new route for future experiments and applications.  For example,  in distinction to spin-glass systems, here the non-ergodic metastable phase is characterized by an intriguing state with localized regions of topologically protected patterns, each pattern then forming a potentially noise-resistant memory cell.
Notably, a slight reduction in accuracy enhances memory capacity while preserving a large set of protected structures. This trade-off is particularly advantageous for machine learning purposes, where a larger parameter space allows for encoding more information without requiring perfect accuracy. 

Finally, we note that the TBM addresses the issue of describing ferroelectric behavior dominated by compositional disorder that leads to interacting mesoscopic domains relevant to optical propagation. In this regime, structural analysis and molecular dynamics simulations can prove challenging. \new{Indeed, the TBM is not intended to fully replace detailed microscopic simulations, such as phase-field or ab initio approaches. Rather, it provides a minimal and analytically tractable framework to explore how competing mesoscopic interactions and geometric constraints can produce complex, non-ergodic behavior. The simplicity of the model allows insight into qualitative mechanisms that may remain hidden in fully numerical treatments.} Qualitatively, the TBM isolates key mechanisms that can lead to complex domain phases in functional materials. In line with efforts to formulate minimal yet effective theories that capture essential physics with maximal simplicity \cite{Goodwin2019}, the TBM reveals that it is the competition between effective dipole-dipole-like interactions and the intrinsic two-scale lattice geometry that drives the formation of such intricate domain structures. \new{While KTN:Li is used here as a case study, the TBM is generic and applicable to other polar materials with flux-closure behavior and competing dipolar interactions that manifest superlattices. Future studies could explore applications to, e.g., SBN or multiferroic materials.}



\section{Conclusions}

Concluding, we have introduced and experimentally validated a new explanatory framework for complex behavior observed in disordered ferroelectric crystals that support superlattices, such as KTN. Emergent phenomena are found to be a consequence of the competition between two basic topological defects that emerge from the spontaneous breaking of discrete inversion symmetry and volume-charge screening: domain walls parallel to the principal crystal axes, across which the polarization flips by 180 degrees, and slanted at 45 degrees, across which the polarization abruptly rotates by 90 degrees \cite{Nataf2020, Kittel1949,Landau1935}. Frustration is then triggered by the interplay between the two corresponding spatial scales, each with its specific dipole-dipole interaction, a Heisenberg and a DM-like interaction. The mechanism leads to vortex and metastable phases, depending on the ratio of the two underlying interaction components.  The application of the real-space renormalization group  methods to detect domain patterns reveals the existence of a frustrated percolation phase transition that naturally includes depinning effects, i.e., domain wall reorientation that results in the well-known Barkhausen jumps and slim-loop hysteretic phenomena, when topologically protected states are broken. Our expansion of classical frustration mechanisms, beyond spin-glass and Griffiths effects, opens a new route to explore the intriguing interplay between structure and dynamics and how this can give rise to previously unknown phases, potentially rising to a new level the design and development of innovative miniaturized noise-resistant high-performance information and energy storage devices.


\section*{Acknowledgments}

L. F. and E. DelRe acknowledge support from PNRR MUR Project No. PE0000023-NQSTI, PRIN 2022 MUR Project No. 20223T577Z, Sapienza-Ricerca di Ateneo Projects. P.V. acknowledges the Spanish Ministry of Research and Innovation and Agencia Estatal de Investigación (AEI), MICIN/AEI/10.13039/501100011033, for financial support through Project PID2023-149174NB-I00, funded also by European Regional Development Funds, as well as  Ref. PID2020-113681GB-I00. A. J. A. acknowledges support from the Ministry of Science, Technology and Space (IL Grant No. 4698). P.V. acknowledges the influence of the late Daniel Amit, whose vision and insights underscored the ethical responsibilities of scientific research in the face of armed conflict, systematic violations of human rights, especially regarding the ongoing genocide against the Palestinian people. We thank A. Gabrielli for extremely valuable discussions and suggestions on earlier manuscript versions.

\appendix

\section{Statistical physics of information network diffusion}
\label{StatphysApp}
The Laplacian, $\hat L=\hat D - \hat A$, governs the diffusion dynamics on any regular or heterogeneous structure through the equation
\[ 
\dot{\textbf{s}}(\tau)= - \hat L \textbf{s}(\tau)
\]
 where $\textbf{s}(\tau)$ represents, for instance, the information distribution on the network nodes, $\hat D$ is the degree matrix, and $\hat A$ is the structure's adjacency matrix. The general solution for this equation is $\textbf{s}(\tau)=e^{-\tau\hat L}\textbf{s}(0)$. This lead to the concept of Laplacian density matrix, $\hat \rho(\tau)=\frac{e^{-\tau \hat L}}{Tr (e^{-\tau \hat L})}$, which makes it possible to analyze network structures in greater detail  \cite{LRG,PRR}, opening the door to a thorough analysis through the introduction of the Laplacian entropy, $S(\tau)=-\frac{1}{\log N}\hat \rho (\tau) \log \hat \rho (\tau)$, and the entropic susceptibility (or specific heat). The "canonical" description of heterogeneous networks in analogy with statistical mechanics \cite{LRG, PRR}, is rigorously supported because $\hat L$ plays formally the role of a Hermitian (lower-bounded) Hamiltonian, $Z(\tau)$ is the partition function,  and $\tau$ is a control parameter akin to the inverse temperature. 

In particular, it is the temporal derivative of the entropy which has been denominated the heat capacity of a network \cite{PRR},
\vspace{-1.5ex}
\[ 
    C(\tau)=-\frac{dS}{d(\log \tau)},
\]
as the natural counterpart of the specific heat in classical statistical mechanics. In particular, $C$ is a detector of structural transition points corresponding to the intrinsic characteristic diffusion scales of the network \cite{LRG,PRR}. Indeed, a pronounced peak of C defines $\tau=\tau^*$ and reveals the starting point of a strong deceleration of the information diffusion, separating regions sharing a rather homogeneous distribution of information from the rest of the network. Finally, in the continuum approximation of the Laplacian spectrum, it has been recently proved that the exponent $\gamma$ satisfies the following relation \cite{LRG,PRL-SI}:
\[
C_1=\frac{\Gamma(\gamma+2)}{\Gamma(\gamma+1)}=\gamma+1=\frac{d_s}{2},
\]
where $\Gamma(z)$ is the Euler's gamma function, and the plateau corresponds to half of the spectral dimension, $d_s$, of the network or lattice \cite{PRL-SI}.

\section{Exploring the Metastable Phase}
\label{TempOsc}

\begin{figure}[hbtp]
    \centering
    \includegraphics[width=0.9\columnwidth]{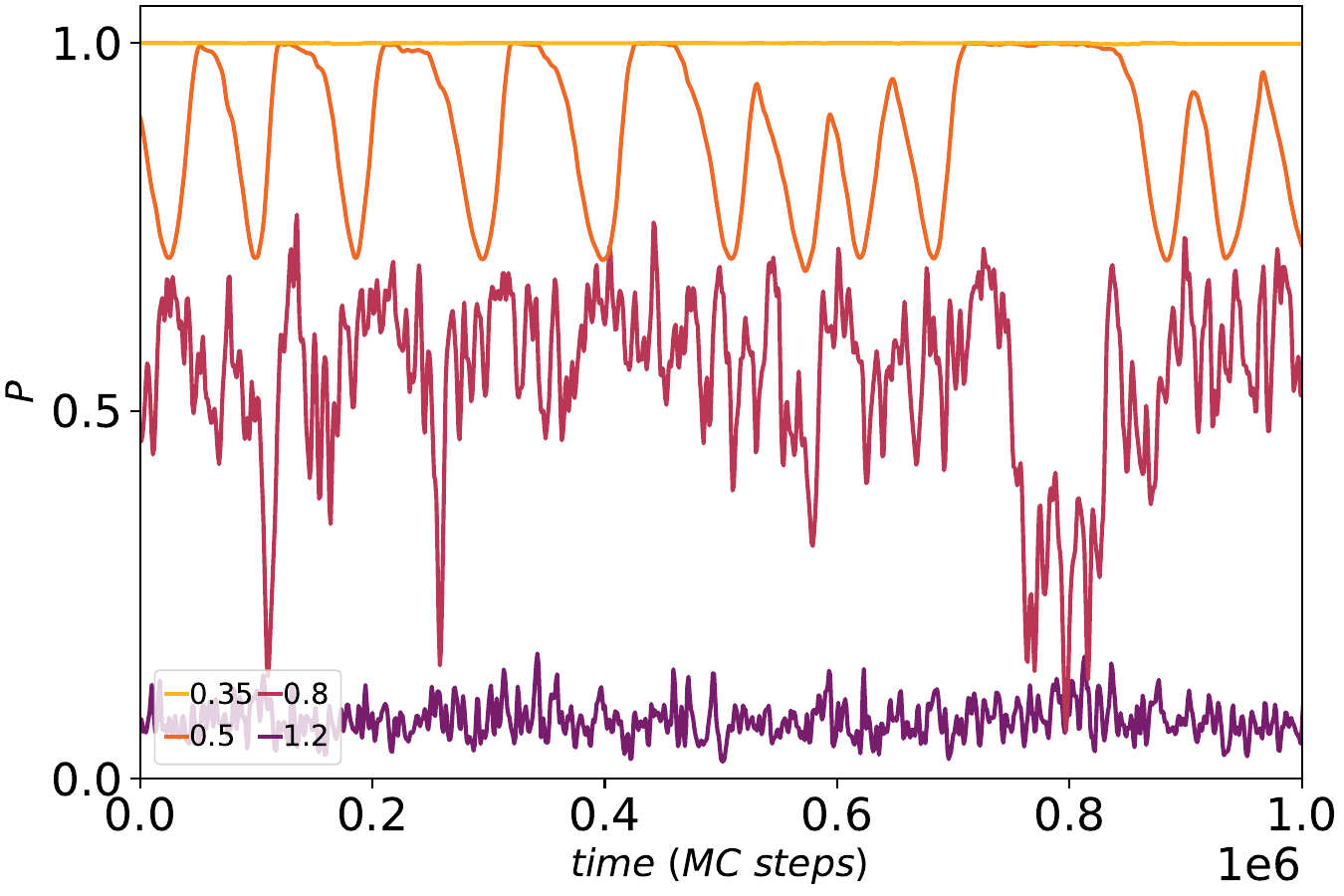}
    \caption{\textbf{Temporal variability.} Polarization order parameter ($P$) as a function of time (in Monte Carlo steps). Different colors represent different values of $T$ (see legend). The broad variability characterizing the metastable phase wildly depends upon the initial conditions for each realization. Parameters: $J/D=1.4$, $L=32$.}
    \label{TempEv}
\end{figure}

For completeness, we present numerical results for the temporal evolution of the polarization order parameter, $P$. Our results reveal the existence of such a metastable regime placed between the vortex and the ferroelectric phase. This is characterized, as shown in Fig. \ref{TempEv}, by broad quasi-periodic temporal oscillations of $P$.
 Anomalously large sampling times would be required to extract good
statistics for the actual mean values and variances. This collective behavior is a straightforward manifestation of the partial ordering of the system, being reminiscent of other types of non-ergodic behavior in structured systems \cite{Villegas2014,Villegas2} but with a completely different origin.
\begin{figure}[hbtp]
    \centering
    \includegraphics[width=1\columnwidth]{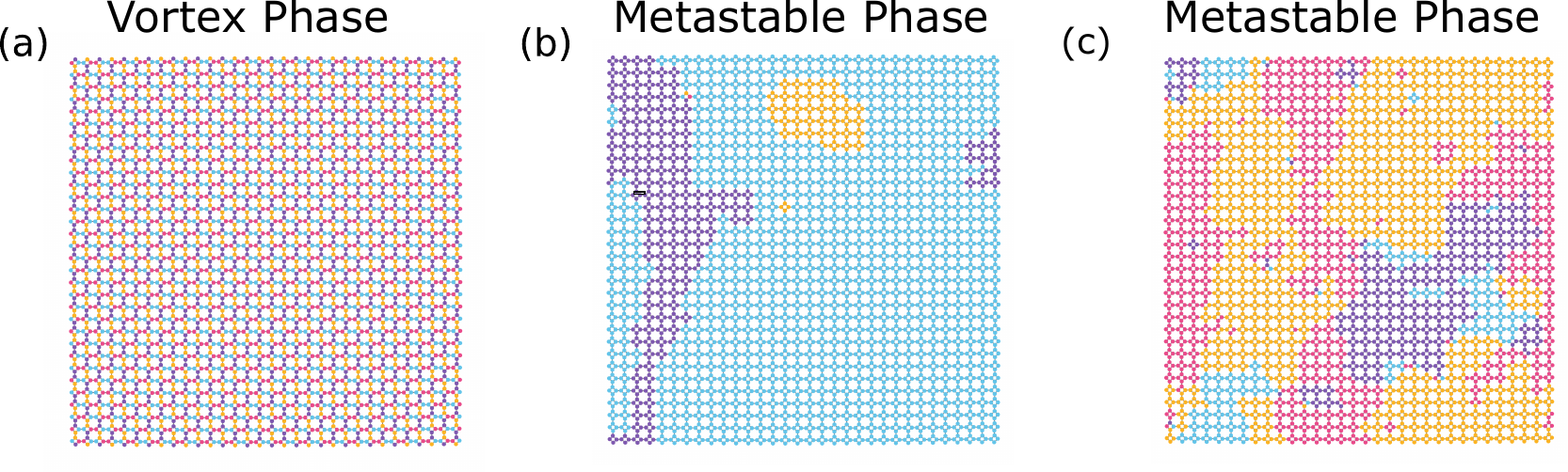}
    \caption{\textbf{Snapshots of several domain arrangements for different phases. (a) $J/D=0.8$, $T=0.35$, (b) $J/D=1.4$, $T=0.5$ and (c) $J/D=1.4$, $T=0.8$. Parameters: $L=32$}.}
    \label{Snapshot}
\end{figure}

We also report different snapshots of the system configurations for multiple phases here to illustrate the emergent order in the two-scale SC lattice. These snapshots have been selected varying the ratio $J/D$ and the temperature $T$. Fig. \ref{Snapshot} (a) illustrates the vortex phase, while Figs. \ref{Snapshot} (b) and (c) show two metastable configurations corresponding to two different temperatures on the phase diagram reported in the previous sections.

\section{Experimental Setup}
\label{ExperimentalApp}
\begin{figure}[hbtp]
    \centering
     \includegraphics[width=0.95\columnwidth]{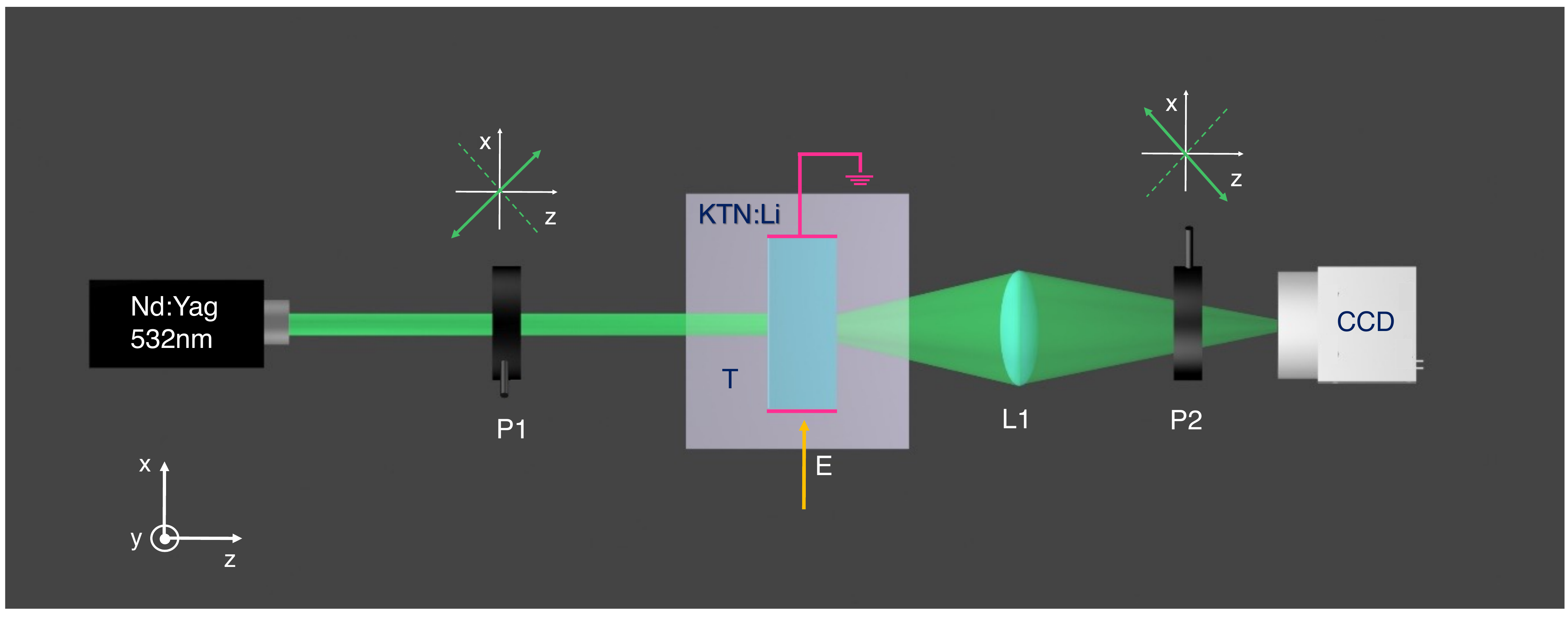}
    \caption{\textbf{Experimental Setup.} Scheme of the cross-polarizer transmission microscopy setup used to resolve mesoscopic polarization textures in KTN:Li.}
    \label{SetUp}
\end{figure}

Laser light from a doubled 30 mW Nd:YAG laser (wavelength $\lambda=2\pi/k_0=532$ nm) is made to propagate along the z-axis through the zero-cut KTN:Li crystal sandwiched in between two crossed polarizers. The sample is biased by a time-constant electric field E along the x-axis. The crystal temperature T is set by a current-controlled Peltier
junction in contact with one of the y facets. Light from the crystal output facet is imaged using a moveable spherical lens L1 (of focal length $50$mm) onto a CMOS camera.
The cross-polarizer technique implemented utilizes the birefringence of polar clusters to alter the polarization state of light, allowing optical observation of microscopic cluster dynamics at micrometer scales. The experimental setup is illustrated in Fig. \ref{SetUp}.

\section{Material}
\label{MaterialApp}
The sample is a zero-cut optical quality polished lithium-enriched solid solution of potassium-tantalate-niobate (KTN: Li) with an average composition $K_{0.964}$Li$_{0.036}$Ta$_{0.60}$Nb$_{0.40}$O$_3$. Its dimensions are 2.33$^{(x)}$ $\times$ 1.96$^{(y)}$ $\times$ 2.03$^{(z)}$ mm. The unit cell manifests random substitutions, a compositional disorder that, as a consequence of the structural flexibility typical of perovskites, leads to locally modified polarizabilities and temperature-dependent nanoscale dipolar structures (nanodisordered ferroelectricity).
The result is a modified ferroelectric behavior dominated by the so-called polar nanoregions (PNRs), characterized by dielectric dispersion and out-of-equilibrium behavior (relaxor ferroelectricity) \cite{Bokov2006}. In our present case, this disorder is itself not homogeneous, manifesting a spatially periodic micrometric oscillation along a specific crystal axis. This is because the sample is grown into a bulk through the top-seeded method, a technique that entails a slight time oscillation in the temperature of the solidifying melt that, in turn, translates into an approximatively periodic 7.5 $\mu$ m pattern along the growth (or pull) axis, which causes in ferroelectrics an equally patterned $T_c$. This pattern conditions the nanoscale dipolar structures that, at the room-temperature Curie point $T_c$= 295K, form an orderly multi-domain protected regular state, the Supercrystal (SC) \cite{Pierangeli2016,Ferraro2017,PercolationSC, LandauSC}.  SCs are thought to originate in a regular lattice of interlocked polarization vortices, a generalization of the spontaneous polarization closed-flux vortex domain pattern dominated by a triangular tiling, compatible with available experimental results and phase-field simulations \cite{LandauSC}. This arrangement effectively screens the polarization charge and elastic stress \cite{Pierangeli2016, LoPresti2020}. 
SC phase also manifests striking optical properties, such as giant broadband refraction (GR) \cite{GR, Falsi2023} and constraint-free wavelength conversion \cite{SHG}.

\section{Percolation analysis of SC structures.}
\label{PerAnalysis}
\begin{center}
\begin{figure}[H]
    \centering
     \includegraphics[width=0.8\columnwidth]{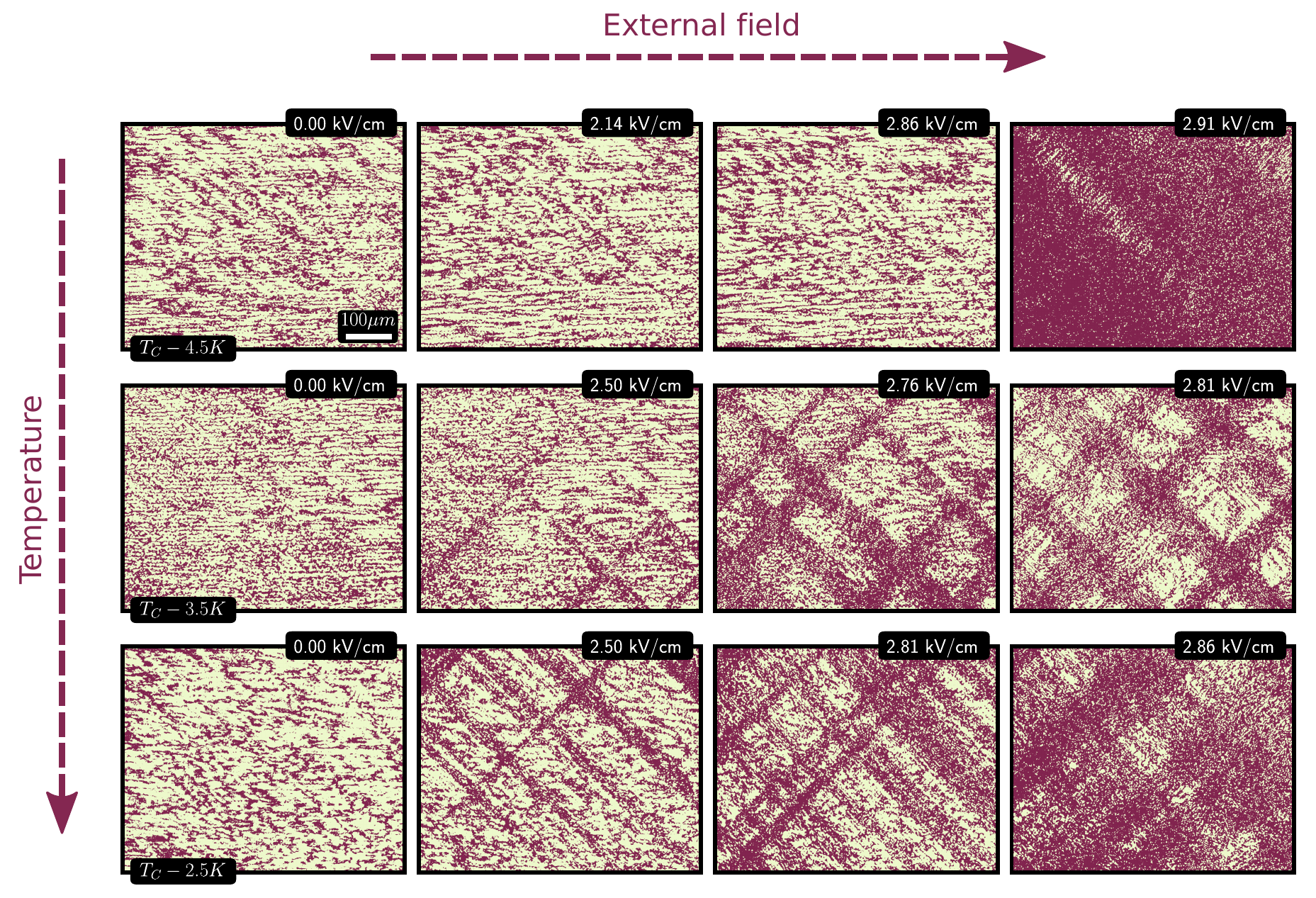}
    \caption{\textbf{Imaging of ferroelectric SC cluster dynamics.} Crossed-polarizer transmission microscopy reveals the temperature- and field-dependent breakdown of the Supercrystal (SC) phase in KTN:Li. The images highlight the emergence of percolative transitions and anisotropic distortions aligned at 45° to the crystal axes, marking the onset of metastable vortex-ferroelectric coexistence predicted by the Topological Breakdown Model.}
    \label{FigExp}
\end{figure}
\end{center}

Basic phenomenology is reported in Fig. \ref{FigExp}, where polarization transmission images through the sample are shown for various temperatures below $T_c=$294K and different external electric fields $E$.
In pure depoled ferroelectrics, random birefringence causes complete depolarization of propagating optical fields, resulting from multiple interferences of randomly scattered waves. In contrast, light propagating in a SC suffers a GR, where waves travel along the principal axes of the crystal without diffracting, remaining fully polarized for a linear polarization along the SC principal axes \cite{Ferraro2017}. The result is that the transmitted light at the output will be a checkerboard-like polarization pattern, with alternating orthogonally polarized states (see Fig.\ref{FigExp}, for $E=0$). This naturally amounts to a natural 3D orthographic projection, opening the possibility of observing optically the microscopic details of SC cluster dynamics even though it is taking place on the micrometer scale in a full 3D volume \cite{PercolationSC}. 
Supporting the main text, in Fig. \ref{FigExp}, a more detailed phenomenology is reported, where polarization transmission images through the sample are shown for various temperatures below $T_c=$294K and different external electric fields $E$.
 The sudden transition at $T_c-$4.5 K evolves into a much richer phenomenology, fully compatible with our hypothesis of an emergent metastable phase.
In fact, for a temperature up $T_C-3.5K$, the SC does not show any more sudden polarization changes. Crossed-polarizer transmission starts to decrease along particular directions, oriented at 45 degrees relative to the crystal principal axes ($E_c$ = 2.81 kV/cm). Despite this, light transmission remains evident in the overall image.
Increasing temperature further ($T_C-2.5K$), SC distortions appear at a lower field, $E_c$=2.5 kV/cm along the inclined paths, analogously to the previous case. As the bias field increases, the dark regions expand along these specific directions until they cover the entire transmitted image at $E_c$= 2.96 kV/cm, analogously to the case at $T_C-4.5K$.

\onecolumngrid
\twocolumngrid

%

\clearpage
\includepdf[pages={1}]{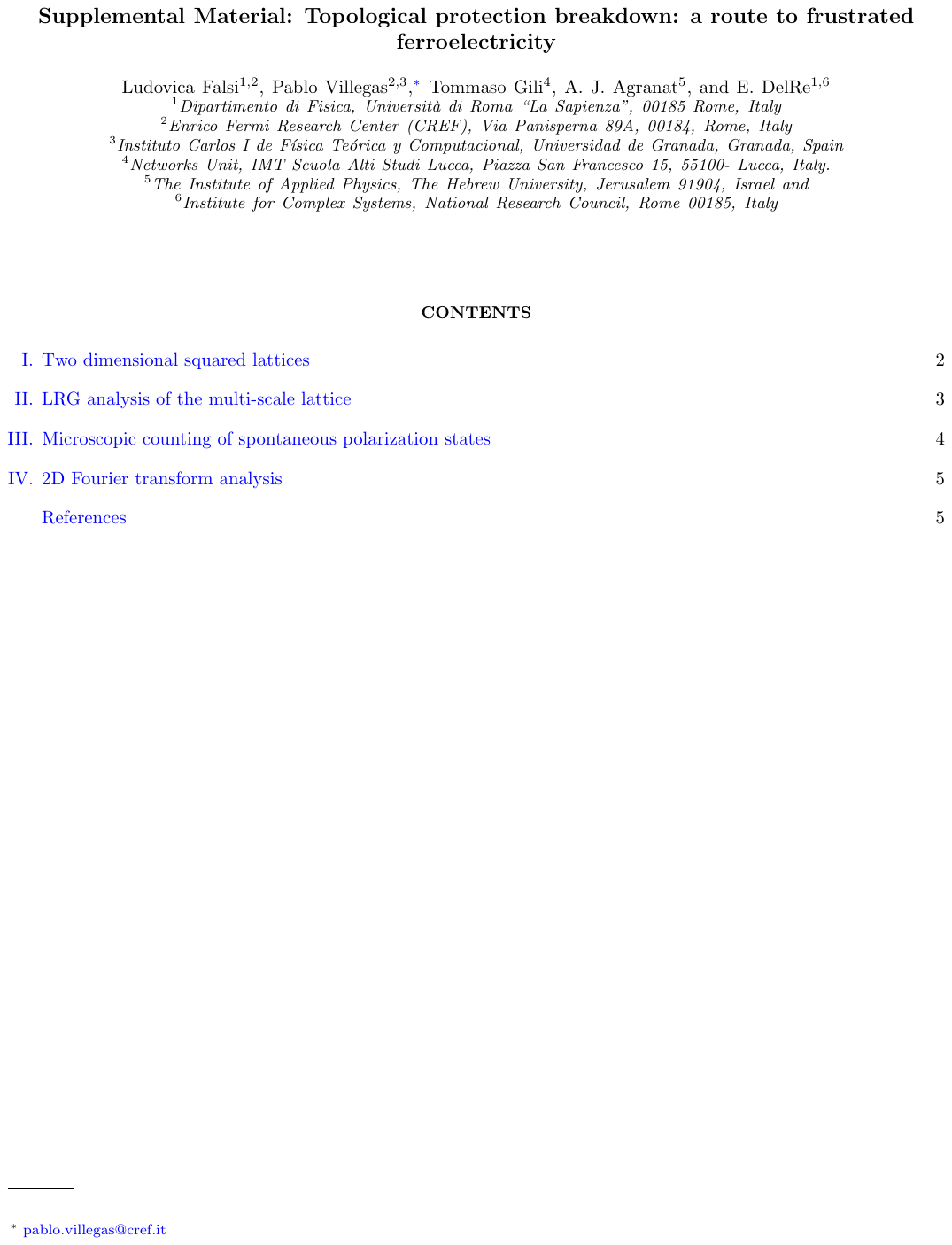}
\clearpage
\includepdf[pages={2}]{Supercrystal_Vortex_Supplementary.pdf}
\clearpage
\includepdf[pages={3}]{Supercrystal_Vortex_Supplementary.pdf}
\clearpage
\includepdf[pages={4}]{Supercrystal_Vortex_Supplementary.pdf}
\clearpage
\includepdf[pages={5}]{Supercrystal_Vortex_Supplementary.pdf}
\end{document}